\documentclass[vecphys]{svmult}

\usepackage{makeidx}         % allows index generation
\usepackage{graphicx}        % standard LaTeX graphics tool
\usepackage{multicol}        % used for the two-column index
\usepackage[bottom]{footmisc}% places footnotes at page bottom
\makeindex             % used for the subject index
\begin{document}

\title*{From ESPRESSO to CODEX}
\author{J.~Liske\inst{1}\and L.~Pasquini\inst{1}\and
P.~Bonifacio\inst{2}\and F.~Bouchy\inst{3}\and
R.F.~Carswell\inst{4}\and S.~Cristiani\inst{2}\and
M.~Dessauges\inst{5}\and S.~D'Odorico\inst{1}\and
V.~D'Odorico\inst{2}\and A.~Grazian\inst{6}\and R.~Garcia-Lopez\inst{7}\and 
M.~Haehnelt\inst{4}\and G.~Israelian\inst{7}\and C.~Lovis\inst{5}\and
E.~Martin\inst{7}\and M.~Mayor\inst{5}\and P.~Molaro\inst{2}\and
M.T.~Murphy\inst{8}\and F.~Pepe\inst{5}\and D.~Queloz\inst{5}\and
R.~Rebolo\inst{7}\and S.~Udry\inst{5}\and E.~Vanzella\inst{2}\and
M.~Viel\inst{2}\and T.~Wiklind\inst{9}\and M.~Zapatero\inst{7}\and 
S.~Zucker\inst{10}}
\institute{$^1$ESO, Karl-Schwarzschild-Str.~2, 85748
Garching, Germany\\
$^2$INAF -- Osservatorio Astronomico di Trieste, Via Tiepolo 11, 34131
Trieste, Italy\\
$^3$Laboratoire d'Astrophysique de Marseille, 13013 Marseille, France\\
$^4$IoA, University of Cambridge, Madingley Road, Cambridge CB3 0HA, UK\\
$^5$Observatoire de Gen{\`e}ve, 51 Ch.~des Maillettes, 1290 Sauverny,
Switzerland\\
$^6$INAF -- Roma, via di Frascati 33, 00040 Monteporzio Catone (Roma), Italy\\
$^{7}$Instituto de Astrofisicade Canarias, 38205 La Laguna, Spain\\
$^{8}$Astrophysics, Swinburne University, Hawthorn, VIC 3122, Australia\\
$^{9}$STScI, 3700 San Martin Drive, Baltimore MD 21218, USA\\
$^{10}$Geophysics and Planetary Sciences, Tel Aviv University, 
Tel Aviv 69978, Israel}
\maketitle

\begin{abstract}
CODEX and ESPRESSO are concepts for ultra-stable, high-resolution
spectrographs at the E-ELT and VLT, respectively. Both instruments are
well motivated by distinct sets of science drivers. However, ESPRESSO
will also be a stepping stone towards CODEX both in a scientific as
well as in a technical sense. Here we discuss this role of ESPRESSO
with respect to one of the most exciting CODEX science cases, i.e. the
dynamical determination of the cosmic expansion history.
\end{abstract}

\section{Introduction}
\label{intro}

CODEX (= COsmic Dynamics EXperiment) is a concept for an extremely
stable, high-resolution optical spectrograph for the European
Extremely Large Telescope (E-ELT). The science case for CODEX
encompasses a large range of topics, including the search for
exo-planets down to earth-like masses, primordial nucleosynthesis and
the possible variation of fundamental constants. However, its prime
science driver is the exploration of the universal expansion history
by detecting and measuring the cosmological redshift drift using QSO
absorption lines. This is also one of the 9 `prominent' science cases
chosen by the E-ELT Science Working Group to be among the highlights
of the entire E-ELT science case. A description of the CODEX project
as a whole was given by \cite{Pasquini05}.

The recognition that an ultra-stable, high-efficiency, high-resolution
optical spectrograph would not only be an extremely valuable
instrument for the E-ELT but also for the VLT led to the development
of the ESPRESSO concept (= Echelle Spectrograph for PREcision Super
Stable Observations, see L.~Pasquini's contribution to these
proceedings). Again, there are a large number of applications for such
an instrument, as several other contributions to these proceedings
have highlighted. Hence, there is also a very strong science case for
ESPRESSO, including e.g.\ detailed studies of the intergalactic medium
and of stellar abundances.

However, apart from these scientific drivers, ESPRESSO will also
fulfil another role: CODEX will represent a major development in
high-resolution optical spectrographs compared to existing instruments
such as UVES and HARPS. In order to achieve its science goals CODEX
will have to deliver an exceptional radial velocity accuracy and
stability, i.e.\ $2$~cm~s$^{-1}$ over a timescale of $\sim 20$~yr.
Although the basic design concepts are already in place, several of
the sub-systems needed to achieve the CODEX requirements do not
currently exist. However, they will be implemented in ESPRESSO for the
first time. Hence, in many respects ESPRESSO will be a CODEX precursor
instrument that will allow us to test and gain experience with the
novel aspects of these instruments. Here we will discuss ESPRESSO in
the context of its CODEX precursor role, both in a technical sense as
well as with respect to the main CODEX science driver, which we
briefly describe next.

\section{Cosmic Dynamics}
\begin{figure}
\centering
\includegraphics[height=8cm,angle=-90]{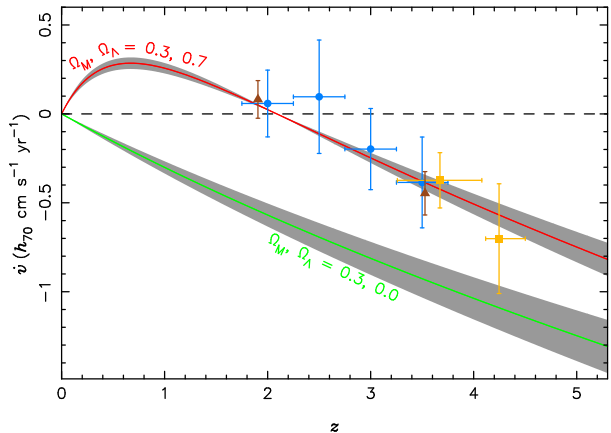}
\caption{The solid lines show the redshift drift as a function of
  redshift in velocity units for two different combinations of
  $\Omega_{\rm M}$ and $\Omega_\Lambda$ as indicated, and a Hubble
  constant of $H_0 = 70$~km~s$^{-1}$~Mpc$^{-1}$. The grey shaded areas
  result from varying $H_0$ by $\pm 8$~km~s$^{-1}$~Mpc$^{-1}$. The
  three sets of `data' points show Monte Carlo simulations of a
  redshift drift experiment with CODEX/E-ELT using a total of
  $4000$~h observing time and a total experiment duration of
  $20$~yr. Each of the sets of points corresponds to a different
  implementation of the drift experiment pursuing different
  observational goals. See \cite{Liske08} for more details.}
\label{vdot}       % Give a unique label
\end{figure}

The discovery from type Ia SNe that the universal Hubble expansion
appears to have begun accelerating at a relatively recent epoch, and
its profound implications for fundamental physics have sparked an
intense interest in the observational exploration of the Universe's
expansion history. Several methods for measuring the Hubble parameter
$H(z)$ already exist but all of them are either geometric in nature or
use the dynamics of localised density perturbations. The simplest,
cleanest and most direct method of determining the expansion history,
however, is to observe the dynamics of the global Robertson-Walker
metric itself. One way to achieve this is by measuring the so-called
redshift drift, i.e.\ the tiny, systematic drift as a function of time
in the redshifts of cosmologically distant sources (Fig.\
\ref{vdot}). This effect is directly caused by the de- or acceleration
of the universal expansion and can hence be used to determine its
history. A measurement of this effect would be able to provide
evidence of the acceleration that is entirely independent of the SNIa
results or any other cosmological observations, and that does not
require any cosmological or astrophysical assumptions at all. It would
also provide $H(z)$ measurements over a redshift range inaccessible by
other methods. Recently, \cite{Liske08} found that a $42$-m telescope
would indeed have the photon collecting power to detect the redshift
drift by monitoring the redshifts of QSO absorption lines over a
timescale of $\sim 20$ years (Fig.\ \ref{vdot}), providing a strong
motivation for a CODEX-like instrument at the E-ELT.

\section{ESPRESSO as a CODEX Pathfinder}
\subsection{Technical aspects}
The UVES and HARPS experiences have allowed us to identify a number of
properties that a spectrograph must feature in order to deliver
exceptional radial velocity accuracy and long-term stability,
including: simultaneous wavelength calibration, a fully passive
instrument with zero human access, located inside a vacuum tank which
is itself located inside a nested environment in an underground
facility that allows progressively more precise temperature and
pressure control, and the flux-weighted timing of observations with
sub-second precision. Temperature control of the CCD will be
particularly important, while high system throughput will also be of
the essence. Some of these concepts are already established. However,
two of the most critical aspects are also those requiring the most
R\&D: light scrambling and wavelength calibration.

At a resolution of $150\,000$ a typical pointing accuracy of $\sim
0.05$~arcsec corresponds to an error of $100$~m~s$^{-1}$ necessitating a
scrambling gain of $\sim 5000$ in order to reach $2$~cm~s$^{-1}$
accuracy. Hence, in addition to any fibre we will require a dedicated
scrambling device to ensure that a photon's position on the CCD only
depends on its wavelength but not on its position on the entrance
aperture.

Current wavelength calibration sources such as ThAr lamps and I$_2$
cells are sub-optimal in several respects, their non-uniformity and
lack of long-term stability being among the concerns. However, a new
concept for wavelength calibration has recently emerged. A `laser
frequency comb' system provides a series of uniformly spaced, very
narrow lines whose absolute positions are known a priori with a
relative precision of $\sim$$10^{-12}$ (see A.~Manescau's
contribution; \cite{Murphy07}).

Neither of the two systems above currently exist, but they would be
developed for ESPRESSO. Being able to test and validate them `on the
sky' would provide valuable experience and input for further
improvements.

\subsection{Scientific aspects}
The scientific goals of CODEX are sufficiently removed from current
observational reality to require validation and demonstration of
feasibility of all aspects of data handling and analysis. This
includes data acquisition strategies (e.g.\ minimum and maximum viable
exposure times), the tracking of CCD distortions, and the accuracy of
flat-fielding, sky subtraction and scattered light corrections. 
The extraction of the cosmological signal from the data will also
require testing. Issues include how to deal with QSO variability and
the accuracy of the conversion to the cosmological reference frame. In
addition ESPRESSO will allow us to determine the currently unknown
intrinsic widths of the narrowest metal absorption lines in order to
reassess their usefulness for the drift experiment, and to look for
any sources of astrophysical noise so far overlooked.

Data on QSOs collected with ESPRESSO for other scientific purposes
would allow us to address all of the above issues. We estimate that
$\sim 200$ hours of observations of the brightest known QSOs would
provide an end-to-end system verification, from data acquisition to
signal extraction, at the level of $\sim 30$--$40$~cm~s$^{-1}$.

\begin{figure}
\centering
\includegraphics[height=8cm,angle=-90]{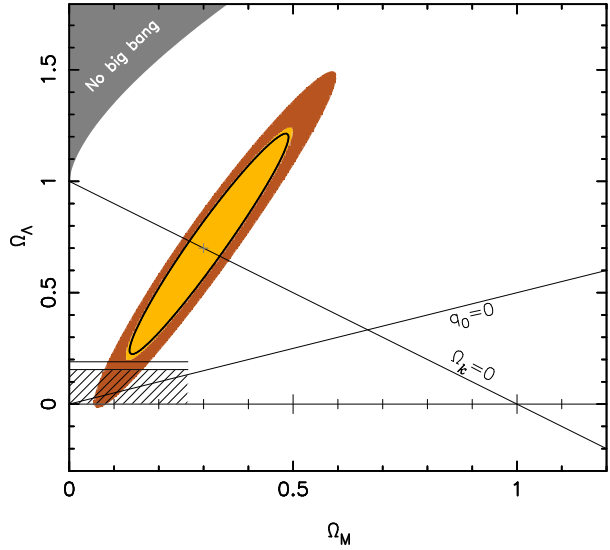}
\caption{Expected constraints in the $\Omega_\Lambda$-$\Omega_{\rm M}$
  plane from the redshift drift experiment with CODEX/E-ELT described
  in detail in \cite{Liske08}. The coloured ellipses show the joint
  $68$ and $90$ per cent confidence regions that result from a total
  integration time of $4000$~h and a total experiment duration of
  $\Delta t = 20$~yr. The hashed region indicates the $95$ per cent
  lower limit on $\Omega_\Lambda$. The solid contour shows the
  improvement of the $68$ per cent confidence region that results from
  the additional investment of $4000$~h of observing time using
  ESPRESSO on the VLT in its `SuperHarps' mode (i.e.\ at $R \approx
  150\,000$ and using one UT), assuming that these observations take
  place $\sim 8$~yr before the start of the CODEX observations. The
  $95$ per cent lower limit on $\Omega_\Lambda$ of the combined
  experiment is shown has the horizontal line above the shaded
  region. Flat cosmologies and the boundary between current de- and
  acceleration are marked by solid black lines. The dark shaded region
  in the upper left corner designates the regime of `bouncing
  universe' cosmologies which have no big bang in the past.}
\label{ol_om}       % Give a unique label
\end{figure}

\section{CODEX + ESPRESSO = $\dot z$?}
Since ESPRESSO will have characteristics similar to those of CODEX the
question arises whether ESPRESSO can be used to make a start on the
redshift drift experiment. The idea is that since ESPRESSO would be
available several years before CODEX data appropriately collected with
ESPRESSO could serve as a `zeroth' epoch measurement, effectively
extending the time baseline of the experiment for a few years, thereby
improving the final result without delaying it. Fig.\ \ref{ol_om}
shows the comparison of the cosmological constraints in the
$\Omega_\Lambda$-$\Omega_{\rm M}$ plane expected from a drift
experiment using CODEX only (coloured ellipses) and CODEX + ESPRESSO
(solid contour), where we have assumed that ESPRESSO would take up
operation $8$ years before CODEX. Evidently the extension of the time
baseline by $8$ years is not enough to offset the lack of the VLT's
photon collecting power compared to the E-ELT: the improvement of the
constraints is only quite modest, with the lower limit on
$\Omega_\Lambda$ increasing by $\sim 20$ per cent.

%%%%%%%%%%%%%%%%%%%%%%%%%%%%%%%%%%%%%%%%%%%%%%%%%%%%%%%%%%%%%%%%%%%%%%  }

%%%%%%%%%%%%%%%%%%%%%%%%%%%%%%%%%%%%%%%%%%%%%%%%%%%%%%%%%%%%%%%%%%%%%%

\printindex
\end{document}